\documentclass{aa}
\usepackage{graphicx}
\usepackage{multirow}

\begin{document}
\title{The effects of resistivity and viscosity on the Kelvin- \\ Helmholtz instability in oscillating coronal loops}
\titlerunning{The effects of resistivity and viscosity on the KHI in oscillating coronal loops}
\author{T. A. Howson \and I. De Moortel \and P. Antolin}
\institute{School of Mathematics and Statistics, University of St. Andrews, St. Andrews, Fife, KY16 9SS, U.K.}

\abstract{}
{Investigate the effects of resistivity and viscosity on the onset and growth of the Kelvin-Helmholtz instability (KHI) in an oscillating coronal loop.}
{We modelled a standing kink wave in a density-enhanced loop with the three dimensional (3-D), resistive magnetohydrodynamics code, Lare3d. We conducted a parameter study on the viscosity and resistivity coefficients to examine the effects of dissipation on the KHI.}  
{Enhancing the viscosity ($\nu$) and resistivity ($\eta$) acts to suppress the KHI. Larger values of $\eta$ and $\nu$ delay the formation of the instability and, in some cases, prevent the onset completely. This leads to the earlier onset of heating for smaller values of the transport coefficients. We note that viscosity has a greater effect on the development of the KHI than resistivity. Furthermore, when using anomalous resistivity, the Ohmic heating rate associated with the KHI may be greater than that associated with the phase mixing that occurs in an instability-suppressed regime (using uniform resistivity).}
{From our study, it is clear that the heating rate crucially depends on the formation of small length scales (influenced by the numerical resolution) as well as the values of resistivity and viscosity. As larger values of the transport coefficients suppress the KHI, the onset of heating is delayed but the heating rate is larger. As increased numerical resolution allows smaller length scales to develop, the heating rate will be higher even for the same values of $\eta$ and $\nu$.}
{}
\maketitle

\section{Introduction}\label{sec:introduction}
Magnetohydrodynamic (MHD) turbulence lies at the intersection of the two broad categories of coronal heating theories; wave heating and reconnection heating. Wave energy, which is thought to be plentiful in the solar atmosphere \citep[see][for a review]{IDM_Review}, may be extracted to drive turbulent plasma flows. This, in turn, may braid the coronal magnetic field leading to reconnection and the dissipation of energy in the form of heat. One possible energy source for this MHD turbulence is the potential for Alfv\'{e}nic wave modes to drive the magnetic Kelvin-Helmholtz instability \citep{HPKHI}. These waves are associated with a significant velocity shear and may be excited by the resonant absorption of transversely oscillating coronal loops (e.g. \citealt{Res_Abs, Res_Abs2} or see \citealt{SSR_Goossens} for a review).    

Studies of transverse waves in the solar corona have observed damping times far in excess of those expected in the near-vacuum conditions of the Sun's atmosphere \citep[e.g.][]{PascoeKinkDamp, GodDamp}. In particular, oscillations of coronal loops that have been interpreted as standing kink waves are seen to decay completely within a few periods. Meanwhile, waves of this nature have attracted attention as potential diagnostic tools in the area of coronal seismology \citep[e.g.][]{BRSeismology,AndSeismology}. For example, \citet{PasSeis_DensCont} found that the decay profile of the kink waves could be used to determine the contrast in density between the loop interior and exterior. This may be useful for constraining numerical models such as the one presented below.

The enhanced damping is widely thought to be caused by the process of resonant absorption through which wave energy is transferred from the kink mode to an Alfv\'{e}nic wave. This efficient transfer is caused by a resonance that will occur at any layer within the loop that has an Alfv\'{e}n speed equal to the phase speed of the transverse oscillation. This process is interesting in the context of coronal heating as it generates the small length scales required for efficient energy dissipation. In addition, it has been shown \citep[e.g.][]{Res_Abs3} that the large velocity shear associated with the localised Alfv\'{e}n wave can induce the onset of the Kelvin-Helmholtz instability (KHI). We note that these authors show that even in the absence of the azimuthal Alfv\'en wave, for example, with a step density profile, the instability can still be excited. This instability causes the deformation of the loop's density profile and generates an energy cascade to smaller and smaller length scales, culminating in a regime in which kinetic energy is expected to be readily deposited as heat. 

Currently, observations of the instability in the solar atmosphere are limited. However, recent prominence observations have been proposed as evidence for its existence. They show thread-like substructure and out-of-phase behaviour between their Doppler velocities and their plane-of-the-sky transverse displacement. These were interpreted as the manifestation of resonant absorption and the turbulent aftermath of the KHI \citep{KHI_strands, KHI_strands2}. In addition, the observed decayless oscillation of kink waves \citep{Anfino2013, Anfino2015} may be explained by this combination \citep{Loop_compression}. Furthermore, using data from the Solar Dynamics Observatory, \citet{Foullon_KHI_inCMEs} and \citet{Ofman_KHI_inCMEs} have interpreted the formation of vortices at the boundary of coronal mass ejections as evidence of the KHI. However, in these cases, the instability is triggered by high speed, field-aligned flows rather than transverse oscillations.

Analytic studies of the KHI in magnetised plasmas have been conducted \citep[e.g.][]{BPKHI}, however a full consideration of non-linear waves in a dissipative medium can only be achieved numerically. In plasmas, the presence of a magnetic field aligned parallel to a velocity shear stabilises the KHI via the action of the magnetic tension force \citep{KHI_Stable}. However, this inhibiting effect may not be as strong as previously thought, particularly if the resonance layer is thin or the field is only weakly twisted \citep{Cowling_MHD, Twist_OK}. Furthermore, it is understood that viscous and resistive effects also suppress the growth and, in some cases, the formation of the instability \citep{PatrickKHI}. For example, viscous forces act to limit the velocity shear responsible for the KHI. As a consequence, it is essential that numerical models have sufficiently high spatial resolution to ensure that artificial viscous effects do not restrict the onset of the instability. 

It is widely anticipated that the values of the Lundquist and Reynolds numbers in the solar corona are far in excess of those currently achievable in numerical simulations. Therefore, since the nature of energy dissipation in the corona cannot be replicated correctly, we aim to evaluate the effects of enhanced transport coefficients on the development of the KHI. Previous studies \citep[e.g.][]{Res_Abs_Eta1, Res_Abs_Eta2} have considered the effects of viscosity and resistivity on the process of resonant absorption and the associated heating. Typically, only with the use of dissipation coefficients enhanced by many orders of magnitude above coronal levels, is significant heating achieved on the cooling timescale \citep{IDMPC}. The small length scales associated with the formation of the KHI have been invoked as a possible explanation for localised heating.

Investigations of the magnetic reconnection associated with the instability \citep{KHI_Reconnection} have shown that the onset of the KHI enhances the rate of energy dissipation. Following the onset of the instability, \citet{PatrickKHI} showed that intricate current sheets form in a manner not too dissimilar to the well-known Parker braiding regime \citep{Parker_Braiding}. Unfortunately, the turbulent conditions that form are not properly resolved in full, 3-D MHD simulations. Indeed, \citet{KHI_strands2} showed that poor numerical resolution will have a significant effect on the formation and evolution of KHI vortices. Low spatial resolution is associated with artificially enhanced dissipation that causes a loss of energy from the domain.

In this paper, we present the results of a parameter study that investigates the effects of resistivity and viscosity on the development of the KHI. In Section 2, we present the numerical model followed by a description of the results in Section 3. A discussion and our conclusions are presented in Section 4.

\section{Numerical method}

\subsection{Initial set-up}

\begin{figure}[h]
  \centering
  \includegraphics[width=0.5\textwidth]{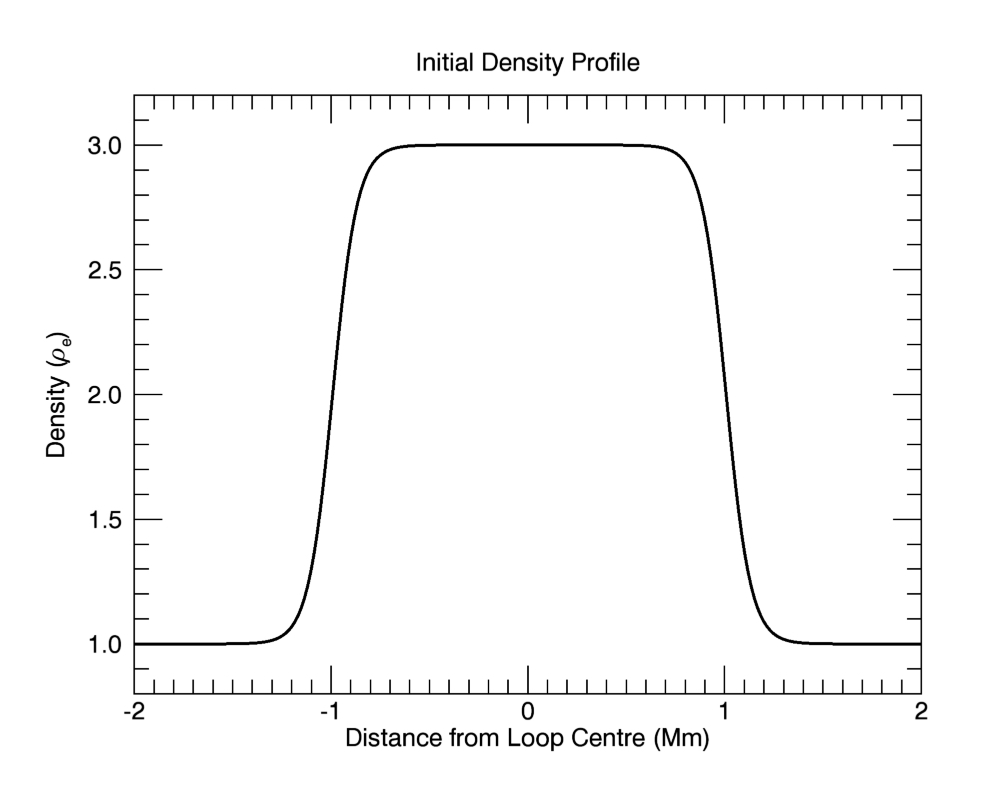}
  \caption{Initial density profile through the cross-section of the loop. The density is normalised to the initial exterior density, $\rho_e = 8.4 \times 10^{-13}$ kg m$^{-3} $.}
  \label{in_dens}
\end{figure}

\begin{figure}[h]
  \centering
  \includegraphics[width=0.5\textwidth]{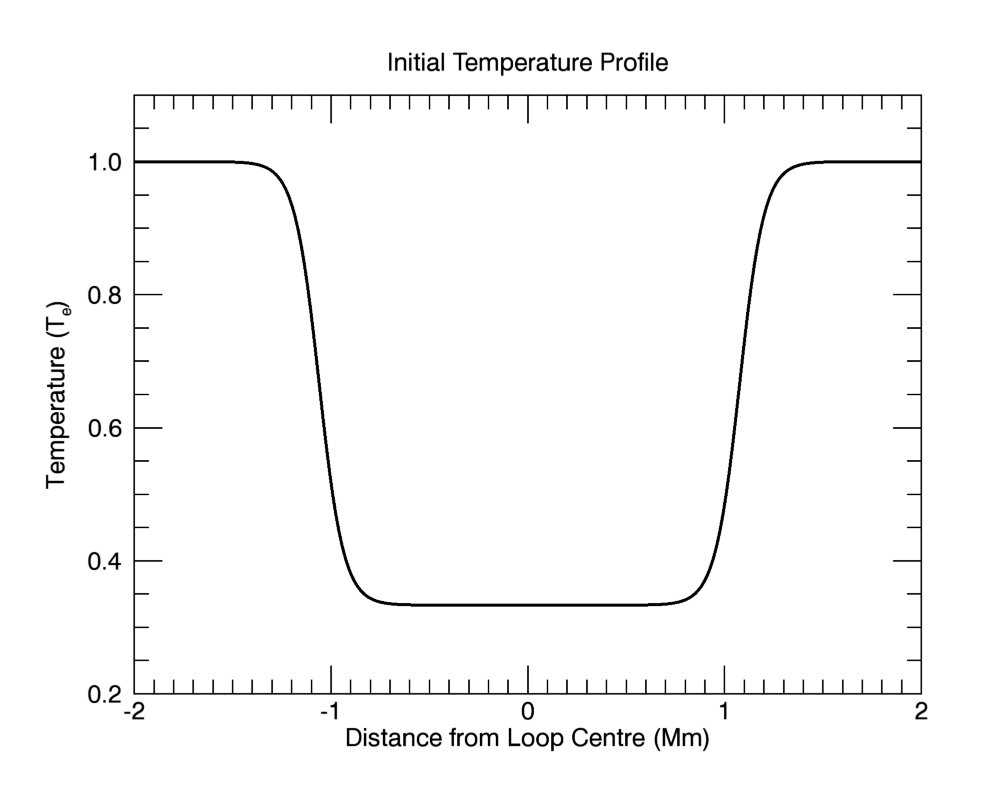}
  \caption{Initial temperature profile through the cross-section of the loop. Here we have normalised the temperature to the initial exterior temperature, $T_e = 2.5$ MK.}
  \label{in_temp}
\end{figure}

As in \citet{PatrickKHI}, we modelled a coronal loop as a straight, density-enhanced flux tube. The loop had length 200 Mm and a radius of approximately 1 Mm. The loop cross-section exhibited a density profile as shown in Figure \ref{in_dens}. The density followed a tanh profile and increased from an external level of $\rho_e = 8.4 \times 10^{-13}$ kg m$^{-3}$ to an internal level of $\rho_i = 3 \rho_e = 2.52 \times 10^{-12}$ kg m$^{-3}$. We labelled the central region with near uniform density as the core and the transition between the high density interior and low density exterior as the shell region. A uniform magnetic field aligned with the loop axis ($y$ direction) and of strength 21 G was included in the domain. The coronal loop is maintained in pressure balance by the temperature being set equal to

\begin{equation} T = \frac{P_0}{\rho}, \end{equation}    

where $\rho$ is the density and $P_0$ is a constant set to ensure the $\text{plasma-}\beta = 0.05$ everywhere. This temperature profile is shown in Figure \ref{in_temp}. In the absence of any initial perturbation, this configuration is maintained in equilibrium. In all of our simulations, we neglected any effects of loop curvature, gravity, thermal conduction and radiation.

At the start of each simulation, we induced the fundamental standing kink mode in the loop with a velocity profile given by

\begin{equation*} v_x = A (\rho - \rho_e) \cos\left(\frac{2 \pi y}{L}\right), \end{equation*} 
\begin{equation} v_y = 0, \end{equation}
\begin{equation*} v_z = 0, \end{equation*}

where $L = 200$ Mm is the length of the loop and $A$ is a constant resulting in the amplitude at the loop apex being 8.3 km s$^{-1}$. This corresponds to an initial velocity of approximately $1 \% $ of the local Alfv\'en speed. Since outside the loop $\rho = \rho_e$, we only perturbed the loop itself and the dense centre experienced the greatest initial velocity. At all times, the footpoints of the loop at $y= \pm 50 $ are fixed with zero velocity. All other variables had zero gradients at the boundaries.

\subsection{Numerical code and domain}

For our numerical experiments, we used the Lagrangian-remap code, Lare3D \citep{Larey}. Using a staggered grid, this code advances the full, 3-D, resistive MHD equations in normalised form that are given by

\begin{equation}\frac{\text{D}\rho}{\text{D}t} = -\rho \vec{\nabla} \cdot \vec{v}, \end{equation}
\begin{equation}\rho \frac{{\text{D}\vec{v}}}{{\text{D}t}} = \vec{J} \times \vec{B} - \vec{\nabla} p + \vec{F}_\nu, \end{equation}
\begin{equation}\rho \frac{{\text{D}\epsilon}}{{\text{D}t}} = \eta |\vec{J}|^2 - p(\vec{\nabla} \cdot \vec{V}) + H_\nu, \end{equation}
\begin{equation}\frac{\text{D}\vec{B}}{\text{D}t}=\left(\vec{B} \cdot \vec{\nabla}\right)\vec{v} - \left(\vec{\nabla} \cdot \vec{v} \right) \vec{B} - \vec{\nabla} \times \left(\eta \vec{\nabla} \times \vec{B}\right). \end{equation}

In this set of equations, all variables have their usual meanings, $\vec{F}_\nu$ is the contribution of viscous forces and $H_\nu$ is the associated viscous heating. These terms are defined to be

\begin{equation*}\vec{F}_\nu = \nu \left(\nabla^2 \vec{v} + \frac{1}{3} \nabla \left(\nabla \cdot \vec{v}\right)\right),\end{equation*}
\begin{equation*}H_\nu = \nu \left( \frac{1}{4} \epsilon_{i,j}^2 - \frac{2}{3} \left( \nabla \cdot \vec{v}\right)^2\right),\end{equation*}

where

\begin{equation*} \epsilon_{i,j} = \frac{\partial v_i}{\partial x_j} + \frac{\partial v_j}{\partial x_i}. \end{equation*}

For all simulations, the numerical domain corresponded to dimensions of 64 Mm $\times$ 200 Mm $\times$ 64 Mm. For most of our experiments we used 512 $\times$ 100 $\times$ 512 grid cells, however in two simulations we also considered the effects of resolution by halving the number of grid cells in both the $x$ and $z$ directions. The fine scale dynamics tended to evolve in the $x$-$z$ plane and hence a much finer resolution was used in these directions than in the $y$ direction. In order to minimise boundary effects, the $x$-$z$ plane was non-uniform with resolution becoming less refined away from the centre of the domain. The non-uniform grid has little effect on the simulation as the loop cross-section only oscillated within a uniform central region. The $y$ direction was uniform along the entire length of the loop. The cell resolution in the centre of the domain was 15.9 km $\times$ 2000 km $\times$ 15.9 km (31.8 km $\times$ 2000 km $\times$ 31.8 km in the low resolution cases). The code included user-controlled resistivity and viscosity terms, which were changed for each run in our parameter study.      

We have conducted a parameter study to determine the effects of resistivity and viscosity on the formation and evolution of the Kelvin-Helmholtz instability. In the Lare3d code, the resistivity is normalised to be the inverse of the Lundquist number and the viscosity is normalised as the inverse of the Reynolds number. We investigated the following values of $\eta$ and $\nu$:

\begin{itemize}
\item Resistivity: $\eta = [10^{-3}, 10^{-4}, 10^{-5}, 10^{-6}, 10^{-20}$],
\item Viscosity: $\nu = [10^{-3}, 10^{-4},  10^{-5}, 10^{-6}, 10^{-20}$].
\end{itemize}

Typically in the corona, we expect the Lundquist number to be of the order $10^{12}$ and so we see that most of the values of $\eta$ investigated are artificially large (a consequence of the numerical resolution limitations). We note that both the numerical viscosity and resistivity are significantly larger than $10^{-20}$ and so the experiments using these values represent an ideal simulation, but where the effective, numerical values are of the order $10^{-5}$--$10^{-6}$ (see below). We also note that the dissipative values investigated may be applicable to oscillations in the chromosphere and prominences where the effects of viscosity and resistivity are expected to be more significant due to the plasma only being partially ionised \citep[e.g. Figure 1 in][]{Soler2015}. In addition, we explored the effects of an anomalous resistivity, $\eta^{*}$, which is triggered only in locations exhibiting currents above a critical level.

\section{Results}
\begin{figure*}
  \centering
  \includegraphics[width=\textwidth]{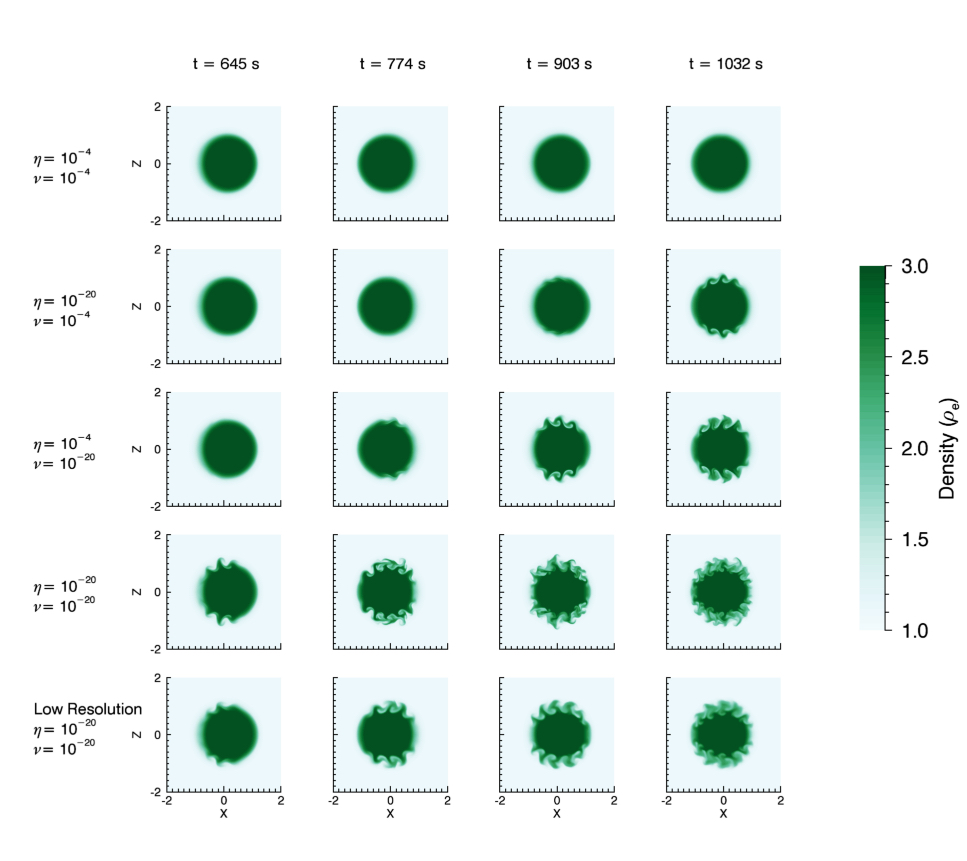}
  \caption{Evolution of the density in the cross-section of the loop apex for different values of resistivity and viscosity. We also include the results of a simulation run at lower spatial resolution. The density is normalised to the initial value of the density outside of the loop, $\rho_e$.}
\label{Dens_Ev}
\end{figure*}

The initial velocity profile excites a transverse oscillation in the form of a standing kink mode. Magnetic tension acts as the restoring force and the loop oscillated with a wave period of approximately 280 s. Within a couple of periods, the resonance between the kink wave and an azimuthal Alfv\'{e}nic wave induces a velocity shear in the shell region of the loop. This process encourages the growth of unstable wave modes that ultimately lead to the development of the Kelvin-Helmholtz instability.

\subsection{Density evolution}

The KHI manifests itself in the deformation of the density profile in the loop's cross-section. The largest effects are observed at the anti-node located at the apex of the loop. Figure \ref{Dens_Ev} highlights the degeneration of the density structure in the shell region for various simulations. We notice that the first unstable mode is the same in each case: we see 3 vortices that rise up first, corresponding to the $m = 3$ mode \citep[see eq. 1 in][]{Terradas2008}. The difference is therefore in the transfer of energy into this mode, which slows down for stronger diffusion. We also see the differences in the presence of higher wavenumbers (e.g. $m = 5$) which are seen in the unsuppressed cases but not in the stronger viscosity cases. For the high resolution runs, since changing the level of viscosity below $10^{-6}$ has no effect on the growth of the instability (also Table \ref{Onset_Tab}), we conclude that the numerical viscosity present in the experiments is of the order $10^{-5}$--$10^{-6}$. We find a similar value for the numerical resistivity and so conclude the numerical Reynolds and Lundquist numbers are approximately $10^5$--$10^6$ as in \citet{PatrickKHI}. Therefore, the bottom two rows correspond to ideal simulations and we identify these with the development of the KHI in an unsuppressed regime. We see from the top row that increasing the viscosity and resistivity by just a couple of orders of magnitude above the numerical level, results in the complete suppression of the instability. Meanwhile, when comparing the two unsuppressed cases, we see that the enhanced numerical dissipation associated with the low resolution simulations limits the small scales that are generated by the KHI.  

The slight asymmetry observed in the loop density in, for example, the first image in the top row, is evidence of the compression of the leading side of the loop during the oscillation \citep{Loop_compression}. Much of this effect is caused by the nature of the initial velocity profile. The core region has a larger velocity than the shell region and so we naturally see a compression at the leading edge of the loop and expansion at the trailing edge. In addition, being denser than the shell region, the loop core has a greater inertia and so a larger magnetic tension force is required to decelerate this plasma. This is a reversible process and an equal and opposite effect occurs half a wave period later. In addition, fluting modes are generated within the oscillating loop and also contribute to the compression of the leading edge \citep{Fluting_modes}.

\begin{figure}[h]
  \centering
  \includegraphics[width=0.5\textwidth]{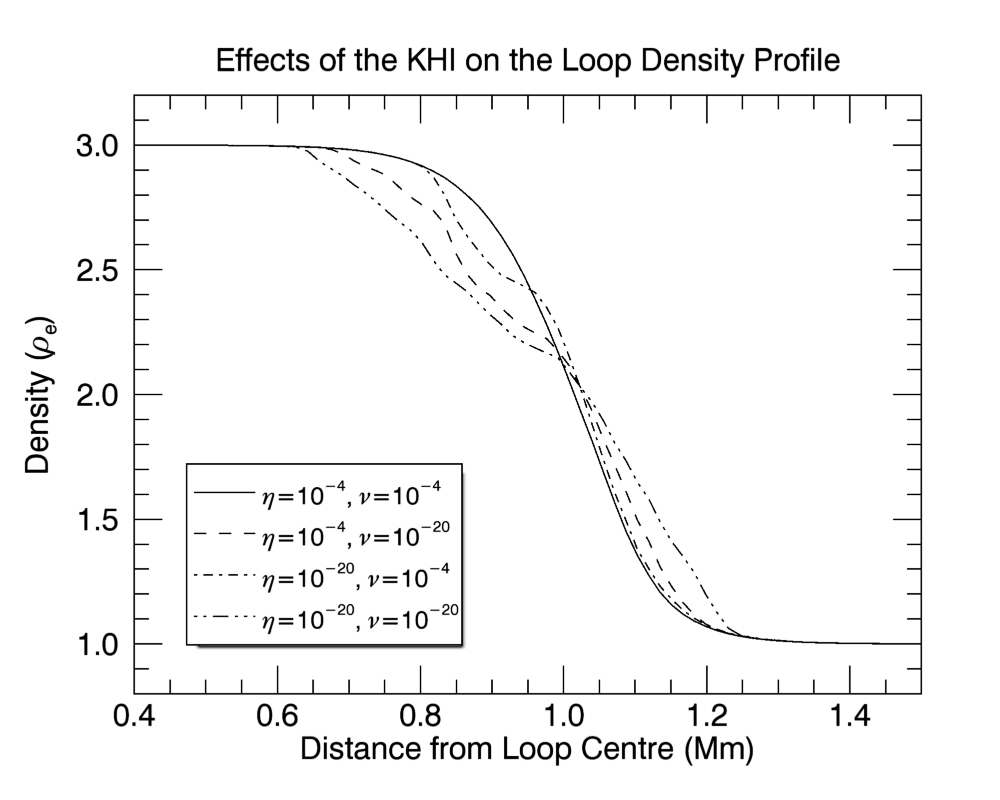}
  \caption{Average density along different radii of the loop apex cross-section at the end of the simulation. The KHI is well-developed in three of the cases and is completely suppressed in the $\left\{\eta=10^{-4}, \nu=10^{-4}\right\}$ simulation.}
  \label{Av_Dens}
\end{figure}

In Figure \ref{Av_Dens}, we examine the effect of the instability on the average density profile of the loop. The plot was produced by taking many different radial cuts of the loop, each forming a different angle with the $x$-axis. We then average these profiles to produce a mean density structure. Since the effect of the KHI is approximately symmetric, we only show half of the loop cross-section. The effect of the instability is to broaden the boundary region and so reduce the average density gradient between the interior and exterior of the loop. We observe the asymmetric effects of viscosity and resistivity in limiting the density deformation. The solid line corresponds to a case where the instability is suppressed completely and so the loop averaged density profile is unchanged from the initial equilibrium (Figure \ref{in_dens}).

In the non-dissipative case $\left\{\eta=10^{-20}, \nu=10^{-20}\right\}$, the instability begins to form after approximately two wave periods. Meanwhile, in the suppressed simulations, it tends to be delayed by some multiple of half the oscillation period (140 s). This is because the rate of energy transfer from the kink mode to the Alfv\'{e}nic mode is not uniform throughout the wave period. Instead, it oscillates with twice the frequency of the global mode. The KHI is most likely to form at times where the energy being injected into the Alfv\'{e}nic wave is large. Hence, it is typically delayed by a factor of half the wave period.

\begin{figure}[h]
  \centering
  \includegraphics[width=0.5\textwidth]{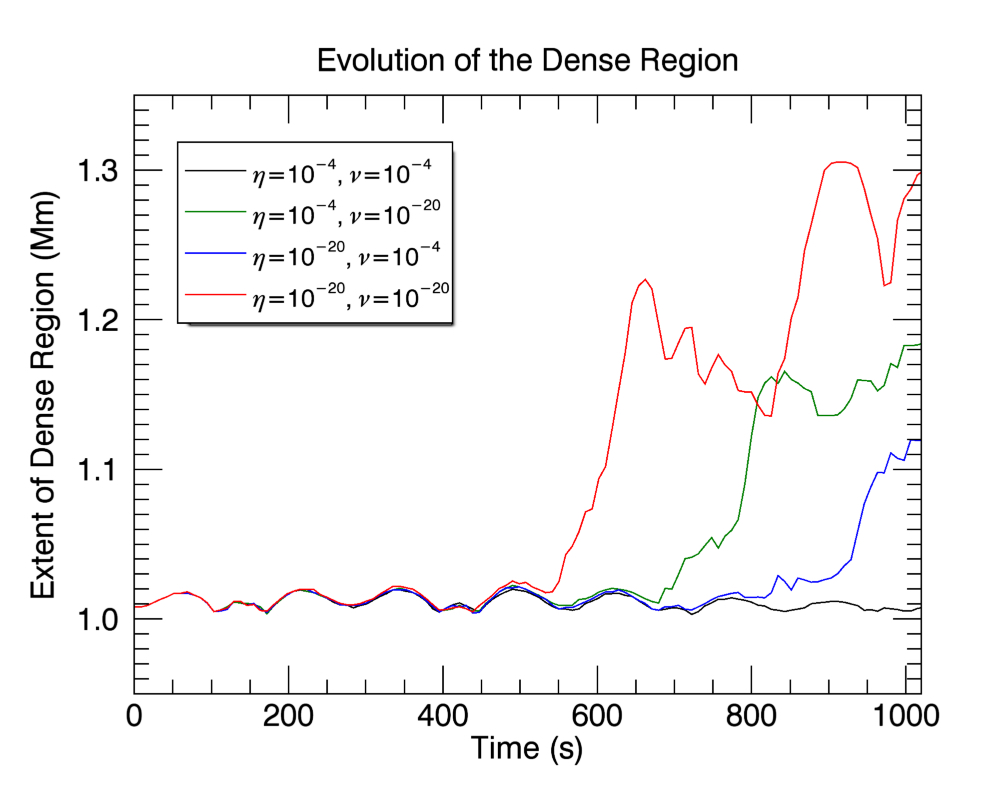}
  \caption{Maximum distance of plasma with density, $\rho = \frac{\rho_e+\rho_i}{2}$ from the loop centre for four different simulations. We use the time of the sharp rise as a measure for the onset time of the KHI (see Table 1).}
  \label{Max_dist}
\end{figure}

In order to quantify the KHI onset time, we monitored the maximum distance of plasma with density equal to $\frac{\rho_e + \rho_i}{2} = 2 $ from the centre of the loop. Initially, this corresponded to a cylindrical shell of plasma located at the midpoint of the boundary layer and is the location of the resonance. Figure \ref{Max_dist} shows the evolution of this maximum distance for four simulations with different levels of KHI suppression.

The rapid rise seen in three of the simulations corresponds to the onset of the KHI. In the experiment for which no sharp rise is observed, the instability did not form. From Figure \ref{Dens_Ev}, we see that as the instability begins, high density plasma moves away from the core of the loop. Identifying the furthest extent of this plasma therefore gives an indication of when the KHI first forms. In addition, the gradient of the initial rise acts as a proxy for the growth rate of the instability. We see that larger transport coefficients act to both delay the onset time of the KHI and slow down subsequent development. 

The small variation prior to the initial rise corresponds to the periodic compression discussed earlier, the compressibility of the Alfv\'{e}nic wave\footnote{We note that the Alfv\'enic motions generated in the boundary layer are not truly incompressible because we are exciting azimuthal modes ($m > 0$)  rather than a purely torsional mode ($m = 0$).} and small errors in tracking the centre of the oscillating loop. The decrease we observe in Figure \ref{Max_dist} for the unsuppressed case at around $t = 600$ s corresponds to the folding over of the vortices (see Figure \ref{Dens_Ev}; row 3, column 3) during the reversal of motion in the kink mode.

\begin{table}[h]
\centering
\label{Onset_Tab}
\begin{tabular}{ccccccc}
\multicolumn{1}{l}{}    & \multicolumn{1}{c}{}            & \multicolumn{5}{c}{$\nu$}                                  \\
\multicolumn{1}{l}{}    & \multicolumn{1}{c|}{}           & $10^{-3}$ & $10^{-4}$ & $10^{-5}$ & $10^{-6}$ & $10^{-20}$ \\ \cline{2-7} 
\multirow{7}{*}{$\eta$} & \multicolumn{1}{c|}{}           &           &     &       &                     &           \\
                        & \multicolumn{1}{c|}{$10^{-3}$}  &           &      &      &                     &       X     \\
                        & \multicolumn{1}{c|}{$10^{-4}$}  &           & X       &   X  &         680            &   680         \\
                        & \multicolumn{1}{c|}{$10^{-5}$}  &           &          &   &                     &          550 \\  
                        & \multicolumn{1}{c|}{$10^{-6}$}  &           &           &  &                     &        540    \\
                        & \multicolumn{1}{c|}{$10^{-20}$}  &    X       &      820  &   550  &         540            &    540       
\end{tabular}
\vspace{1mm}
\caption{Approximate onset times (seconds) of the Kelvin-Helmholtz instability for the completed simulations in the parameter space. The symbol X denotes that the instability was suppressed for the duration of the experiment and a blank entry signifies no data.}
\end{table}

Using this method we identified approximate instability onset times which are displayed in Table \ref{Onset_Tab}. Whilst we see that the viscosity and resistivity do not have a symmetric effect on the onset time, both do contribute to suppressing the KHI. We observe that the viscosity has a greater effect on the formation of the instability. This seems reasonable since, unlike resistivity, the frictional effects of viscosity directly restrict the velocity gradient that triggers the onset of the KHI.

\subsection{Vorticity}

The onset of the Kelvin-Helmholtz instability is associated with the growth of unstable, high wave number Alfv\'{e}nic modes. These waves are particularly sensitive to dissipative forces and are readily suppressed in simulations with high values of $\eta$ and $\nu$. The presence of these wave modes is observed as the development of vortices in the shell region of the oscillating loop. This behaviour enhances the levels of vorticity, $\omega = \vec{\nabla} \times \vec{v}$ present in the domain. Therefore removing energy from these unstable wave modes reduces the levels of plasma vorticity.  

\begin{figure}[h]
  \centering
  \includegraphics[width=0.5\textwidth]{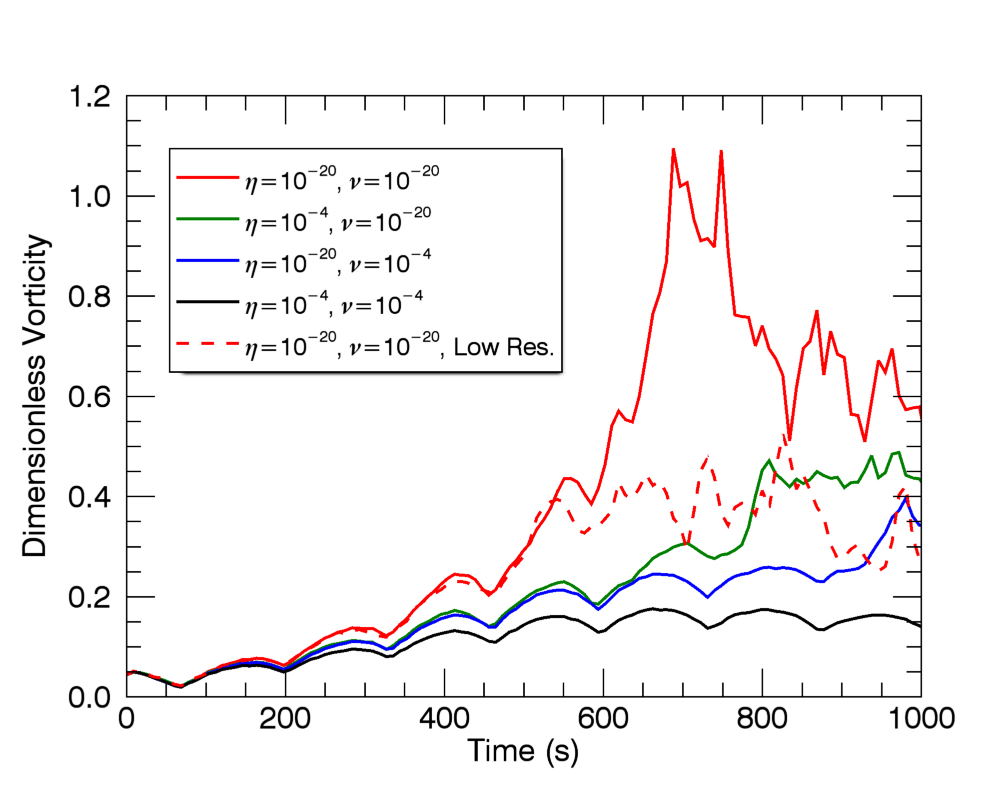}
  \caption{Evolution of the mean of $|\omega_y|$, the component of vorticity perpendicular to the cross-section of the plane, for five simulations.}
  \label{Vort}
\end{figure}

In Figure \ref{Vort}, we examine the evolution of the mean vorticity component perpendicular to the loop cross-section. We consider the effects of both $\eta$ and $\nu$ on the vorticity observed at the loop apex. Since velocities parallel to the loop tend to be small in comparison to those in the loop cross-section, we have $|\omega| \approx |\omega_y|$. Both the azimuthal waves that form through resonant absorption and, to a greater extent, the vortices that form during the growth of the Kelvin-Helmholtz instability are associated with large vorticities. Indeed, we see that in the near-ideal, high resolution case (solid red line), the vorticity reaches a level six times larger than the case in which the instability is suppressed (black line).

\begin{figure*}
  \centering
  \includegraphics[width=\textwidth]{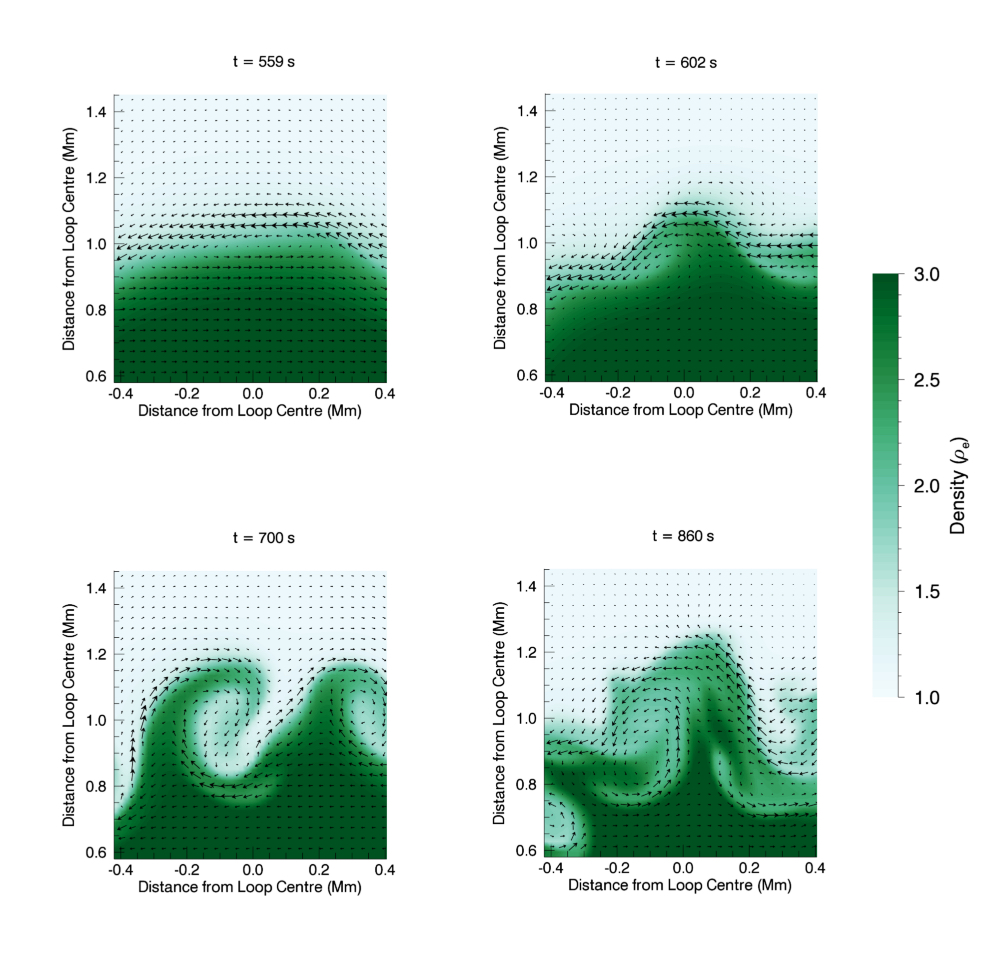}
  \caption{Evolution of the density (green contour plot) and horizontal velocity field (arrows) in the plane of the cross-section at the loop apex. These results are from the approximately ideal simulation with $\left\{\eta = 10^{-20}, \nu=10^{-20}\right\}$. The four snapshots show the plasma characteristics at the boundary of the loop as the KHI develops.}
\label{Dens_Vel}
\end{figure*}

The vorticity minimum observed at around $t=70$ s, is associated with the first displacement maximum in the kink oscillation. At this moment, the standing mode has no velocity associated with it and there has been very little energy transmission to the Alfv\'{e}nic wave in the loop boundary. Following this minimum, we see an increase in vorticity for all cases. This is associated with the flows that form in the loop's shell as a consequence of the resonant absorption. These azimuthal velocities are then subject to phase mixing within the shell of the loop. The Alfv\'{e}nic wave modes have much smaller length scales than the standing kink mode and so velocity gradients, and hence vorticity, tend to be larger than in the initial velocity profile. Further, once resonant absorption begins, Alfv\'{e}nic wave modes are always present in the loop boundary and so no repeat of the vorticity minimum is observed.

Until around $t = 500$ s, the growth in the vorticity shows a periodic pattern in all cases. Approximately every 140 s, we observe a time of steep growth followed by a period of little vorticity change. This periodic behaviour is caused by the mode conversion from the kink wave into Alfv\'{e}nic waves not being constant through each wave period. Energy transfer is more efficient when the kink mode is at maximum displacement and so the vorticity increases fastest at these times. 

\begin{figure*}
  \centering
  \includegraphics[width=\textwidth]{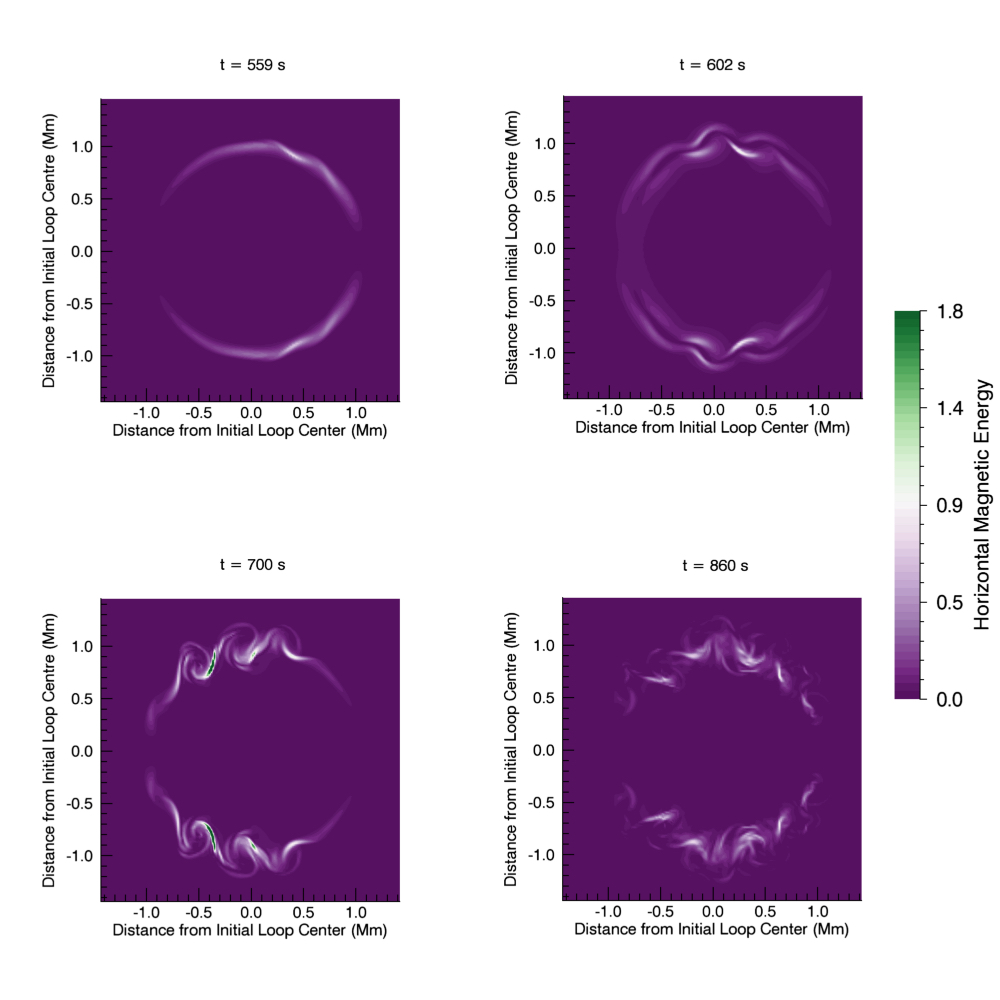}
  \caption{Contour plots of the horizontal magnetic field strength defined as $B_h = \left(B_x^2 + B_z^2\right)^{\frac{1}{2}}$. Here $B_x$ and $B_z$ are the components of the magnetic field in the plane perpendicular to the length of the loop. The simulation shown is the unsuppressed case with $\left\{\eta=10^{-20}, \nu=10^{-20}\right\}$ and the times shown correspond to those in Figure \ref{Dens_Vel}.  }
\label{Mag_KHI}
\end{figure*}

Even after one oscillation, and well before the formation of the KHI in any of the simulations, there are noticeable, $\eta$ and $\nu$-dependent differences in the vorticity. Larger dissipation coefficients have the effect of suppressing the growth of unstable wave modes, and hence the vorticity, present in the domain. These differences are enhanced by the onset of the KHI. This effect is most noticeable when comparing the simulations corresponding to the green and blue lines in Figure \ref{Vort}. Prior to the onset of the instability in either case, the vorticity is remarkably similar. However, once the KHI forms at around $t =  680$ s, the vorticity behaviours diverge. Furthermore, by comparing the solid and dashed red lines, we see that the vorticity present is a function of the spatial resolution. The enhanced numerical dissipation that is present in the low resolution simulation also acts to restrict the vorticity levels.

In order to visualise the increase in vorticity that occurs during the growth of the KHI, in Figure \ref{Dens_Vel} we show the behaviour of the horizontal velocity field for a small section of the loop boundary at four stages in the unsuppressed development of the instability $\left\{\eta=10^{-20}, \nu=10^{-20}\right\}$. The plasma density is also shown to highlight the progression of the KHI. The top-left panel shows the instability as it begins to form. There is a clear velocity gradient across the shell region that is responsible for triggering the instability. In subsequent panels, we notice the formation of the small scale vortices that coincide with the rapid vorticity increase shown in Figure \ref{Vort}.

The deformation of the density profile is associated with strong flows along regions of large density contrast. By the time of the fourth panel, the regions of enhanced density in the shell of the loop have lost their coherent structure. Small length scales have formed in both the density and the velocity field and the simulation is approaching a turbulent regime. Each velocity field arrow corresponds to one grid cell in the numerical domain and we see that the length scales are close to the limit of the simulation's resolution. Beyond this stage, any subsequent development of the instability may become unphysical due to a lack of spatial resolution. 

This observation has repercussions for investigations into the applicability of the KHI as a heating mechanism. Certainly, the small length scales that form during the growth of the instability will enhance the dissipation of wave energy as heat. However, as the energy cascades to smaller scales, simulations will become unable to follow the evolution correctly and energy will be lost from the domain through excessive numerical dissipation. This limitation is a common problem for many models of coronal heating.

The small length scales associated with the turbulent plasma produced by the KHI can also be observed in the evolution of the magnetic energy. The magnetic energy in the domain is dominated by the contribution from the component of the field aligned with the loop axis. Therefore, in order to monitor the magnetic energy associated with the growth of the instability, we define the horizontal magnetic energy as 

\begin{equation} B_h = \sqrt{B_x^2+B_z^2}, \end{equation}

where $B_x$ and $B_z$ are the two horizontal components of the magnetic field. In Figure \ref{Mag_KHI}, we show contour plots of this quantity at the same times as those used in Figure \ref{Dens_Vel}. In the first panel, we observe the concentration of magnetic energy in the boundary of the loop prior to the onset of the KHI. In simulations with a high resistivity, this magnetic energy is dissipated efficiently by the process of phase mixing. This reduces the amount of energy available to drive the instability and therefore suppresses the KHI formation.

In subsequent panels, and in particular at $t = 700 s$, we see the formation of significant spatial gradients in the horizontal magnetic energy. Large values of $B_h$ are associated with the Kelvin-Helmholtz vortices in which localised plasma flows have stressed the magnetic field. The KHI is known to transfer energy from shear flows to compressive flows that encourage energy release through the generation of current sheets and magnetic reconnection (\citealt{KHI_Reconnection}; Fig. 2 in \citealt{PatrickKHI}). The concentration of magnetic field associated with the instability highlights its potential applicability as a coronal heating mechanism. We refer the interested reader to \citet{KHI_Rec1, KHI_Rec2} for studies on magnetic reconnection triggered by the KHI. In simulations where we have non-zero resistivity, the regions of high magnetic energy and, importantly, large field gradients, are subject to significant dissipation and plasma heating. In the fourth panel, we again observe the breakdown of coherent plasma structures as we approach the onset of turbulence. The variation in the horizontal magnetic field strength is apparent in the shell region and is approaching the grid scale of the domain. This further highlights the resolution problem that is encountered when considering the heating properties of the KHI. 

\subsection{Energetics}

\begin{figure}[h]
  \centering
  \includegraphics[width=0.5\textwidth]{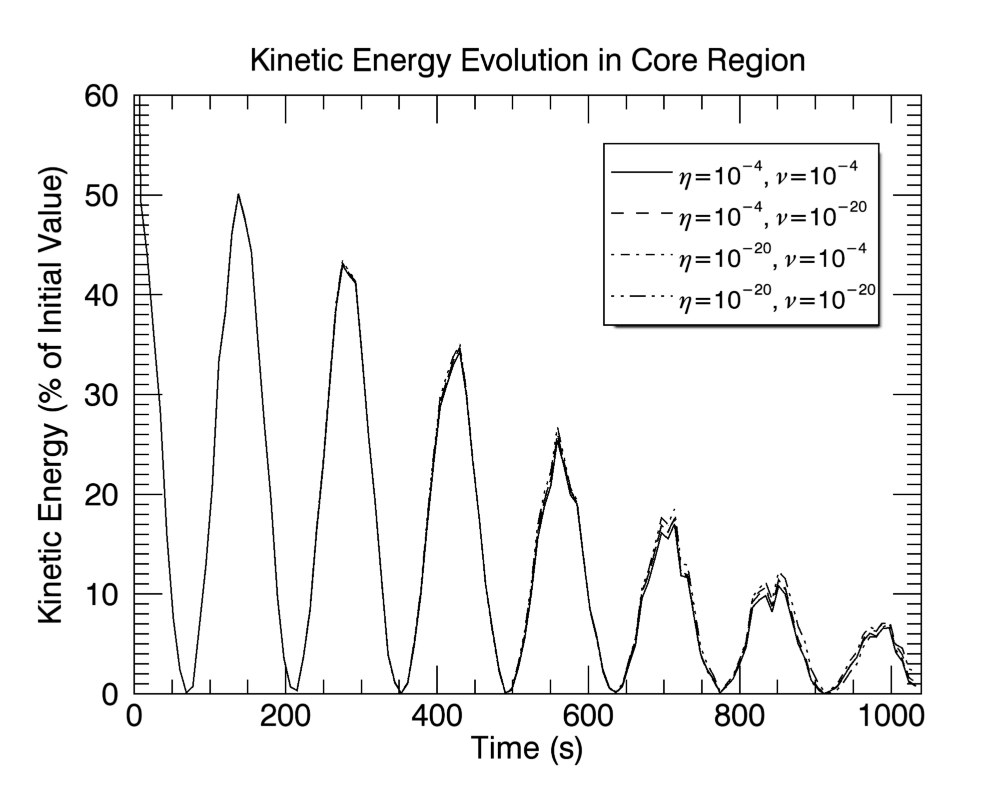}
  \caption{Evolution of the volume integrated kinetic energy in the loop's core region for four simulations.}
  \label{Kin_En_Core}
\end{figure}

In all cases, the kink mode that is excited by the initial velocity profile experiences rapid damping through the well-studied process of resonant absorbtion. This transfers the wave energy from the core region of the loop into the shell. This process is ideal  and so proceeds irrespective of the resistive and viscous effects. Indeed, in Figure \ref{Kin_En_Core}, we see that both the oscillation and loss of kinetic energy from the core are largely independent of the values of $\eta$ and $\nu$. This suggests that the initially excited kink mode is not damped noticeably due to viscous nor resistive effects. Certainly, the initial velocity profile has no horizontal velocity gradient within the loop's core and so we expect the largest effects of viscosity to occur within the boundary layer. It is well established \citep[e.g.][]{Gauss_Damp1, PasSeis_DensCont} that the energy lost from the core region through this process follows a Gaussian damping profile. At later times (after $t=600$ s), small differences in the kinetic energy profiles are noticeable. The near-ideal simulations for which the KHI is strongest show some additional damping. This effect was also noticed by \citet{Fluting_modes}.

\begin{figure}[h]
  \centering
  \includegraphics[width=0.5\textwidth]{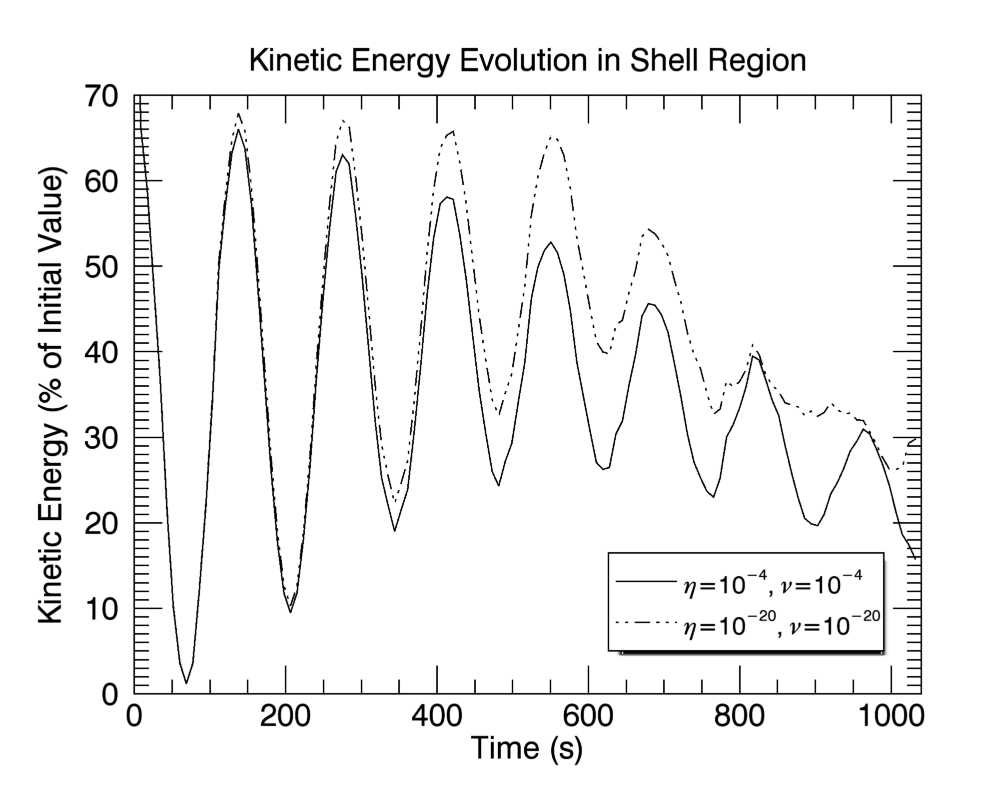}
  \caption{Evolution of the volume integrated kinetic energy in the loop's shell region. The results of two extreme simulations (no suppression for the dotted-dashed line and total suppression for the solid line) are shown. The results of other simulations are bounded by the two curves.}
  \label{Kin_En_Shell}
\end{figure}

Meanwhile, in Figure \ref{Kin_En_Shell}, we monitor the evolution of the kinetic energy within the shell region of the loop. Initially, we see the maxima decay as a function of the dissipative coefficients; larger values of $\eta$ and $\nu$ cause greater Ohmic and viscous heating and so a greater reduction in the kinetic energy. However, even in the approximately ideal case, we see a loss of kinetic energy from the shell. This loss is grid dependent and is enhanced in a lower resolution simulation. We therefore conclude that this energy decrease is due to numerical dissipation and so does not coincide with an increase in thermal energy. At the same time, kinetic energy is constantly being transferred into the shell region from the decaying kink mode. This enhances the power associated with the Alfv\'{e}nic wave, ensuring some part of the loop boundary is in constant motion. This corresponds to the increasing minima seen until $t \approx 600$ s. 

Beyond this point two important things happen; the energy dissipation rate begins to exceed the energy input through resonant absorption and the KHI forms in the non-suppressed case. The former effect is observed as a decrease in wave energy in the shell region and has two causes. Firstly, since the rate of energy input follows a Gaussian profile, as time increases, less energy is being converted from the kink wave mode. Secondly, as the Alfv\'enic mode strengthens, the process of phase mixing becomes more effective at dissipating wave energy. In addition, through the breakdown of the oscillatory pattern, we clearly observe the effects of the KHI in the non-suppressed case in Figure \ref{Kin_En_Shell} beyond $t= 600 $ s. As discussed previously, the formation of the instability has the effect of decreasing the size of typical length scales in both the velocity and the magnetic fields. This amplifies the effects of $\eta$ and $\nu$ and so dramatically increases the rate of wave energy conversion into heat. The small length scales associated with the vortices that form will increase viscous dissipation and the braiding of magnetic field by the turbulent plasma will generate small scale currents and encourage Ohmic dissipation. Unfortunately, the formation of these small scales also increases the magnitude of numerical effects. As a result, even in the approximately ideal case, we observe enhanced kinetic energy loss following the formation of the instability. This is lost from the simulation and not converted into heat. However, it may be tracked more accurately using a locally enhanced resistivity (discussed below).

\begin{figure}[h]
  \centering
  \includegraphics[width=0.5\textwidth]{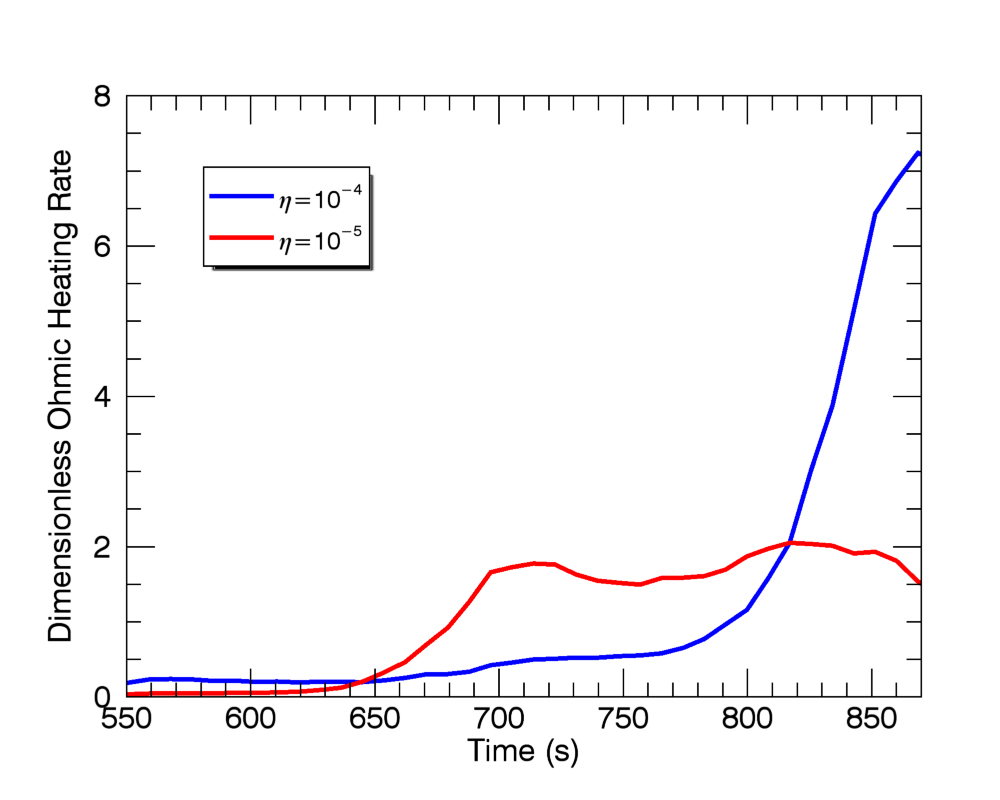}
  \caption{Ohmic heating rate at the loop apex for two simulations in which the KHI forms.}
  \label{Heat_fast}
\end{figure}

The rate of Ohmic heating in a certain location is proportional to $\eta j^2$, where $j =  |\vec{j}|$ is the magnitude of the current. Consequently, in order to obtain comparable Ohmic heating with a resistivity that is an order of magnitude smaller, the size of $j^2$ must be a similar factor larger. In Figure \ref{Heat_fast}, we plot the $\eta j^2$ averaged over the shell region of the loop as a function of time for two different simulations (blue line: $\eta = 10^{-4}$, red line: $\eta = 10^{-5}$). We observe that despite having a lower resistivity value, heating occurs earlier for $\eta = 10^{-5}$ (red line) as the small length scales associated with the KHI develop earlier. For higher values of the resistivity, the development of the instability is somewhat suppressed. However, once it forms, the rate of Ohmic heating is higher (blue line) due to the higher value of the resistivity.

\subsection{Anomalous resistivity}
The turbulent plasma produced by the KHI may have implications for heating the corona. Wave energy is more readily dissipated as heat on the small length scales produced by the formation of vortices. In the turbulent regime, the energy cascade quickly reaches scales on the order of the grid resolution and so cannot be accurately tracked by 3-D MHD simulations. In order to mitigate this problem, it is common practice to use anomalous dissipation coefficients that are triggered if certain criteria are satisfied. In the case of resistivity, this condition is typically dependent on a threshold current. If such currents form, an enhanced resistivity, $\eta^*$ will ensure the associated magnetic energy is dissipated as heat before it is lost from the domain through numerical effects.

\begin{figure}[h]
  \centering
  \includegraphics[width=0.5\textwidth]{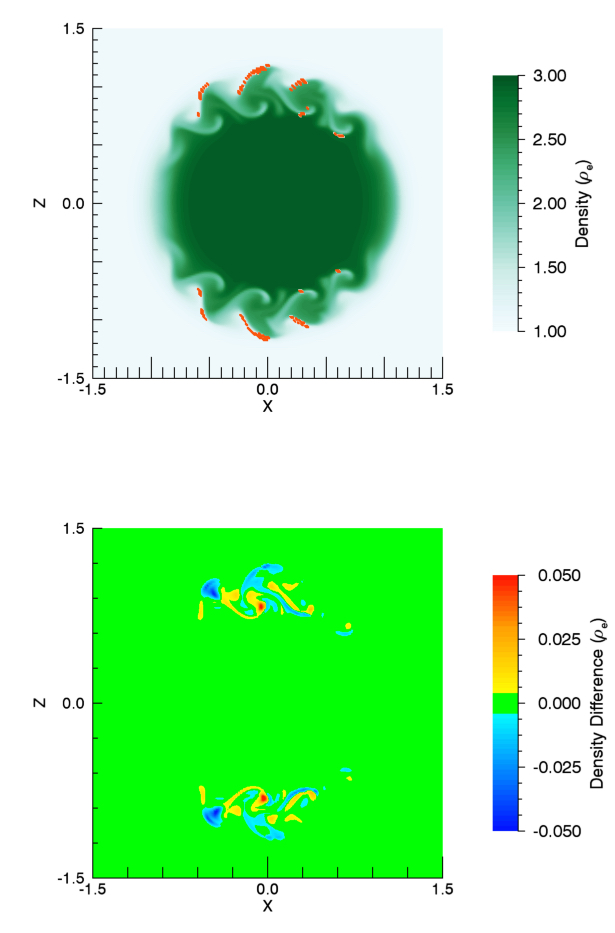}
  \caption{\emph{Upper} - Density profile in the cross-section at the loop apex for the simulation with $\left\{\eta=10^{-20}, \nu=10^{-4}\right\}$ and a critical resistivity, $\eta^*=10^{-4}$.  As in Figure \ref{Dens_Ev}, the density is normalised to the initial exterior density, $\rho_e$. Regions in which the critical resistivity,  is triggered are overplotted in red. \emph{Lower} - Difference plot between the density in the $\left\{\eta=10^{-20}, \nu=10^{-4}\right\}$ simulation and the simulation where $\eta^*$ is considered. Red regions indicate larger densities in the simulation including $\eta^*$ and blue regions indicate larger densities in the original case.}
  \label{crit_cur1}
\end{figure}

The formation of the KHI and the associated turbulence generates small length scales in the magnetic field (see Figure \ref{Mag_KHI}) which are not fully resolved in our simulations. Using a locally enhanced resistivity allows the KHI to develop but limits resolution effects in the turbulent aftermath. Such an approach may be physically valid due to the enhancement of the magnetic Reynolds number that is associated with the development of small length scales.  

In our simulations, the currents are dominated by those induced by the kink mode oscillation near the loop foot points and not by the small scales that form on account of the KHI (see Figure \ref{Mag_KHI}). The instability induced currents are not typically aligned parallel to the magnetic field and so we imposed a critical threshold only on the transverse components of the current. These are greatest at the loop apex where the effects of phase mixing and the KHI are most profound.

\begin{figure}[h]
  \centering
  \includegraphics[width=0.5\textwidth]{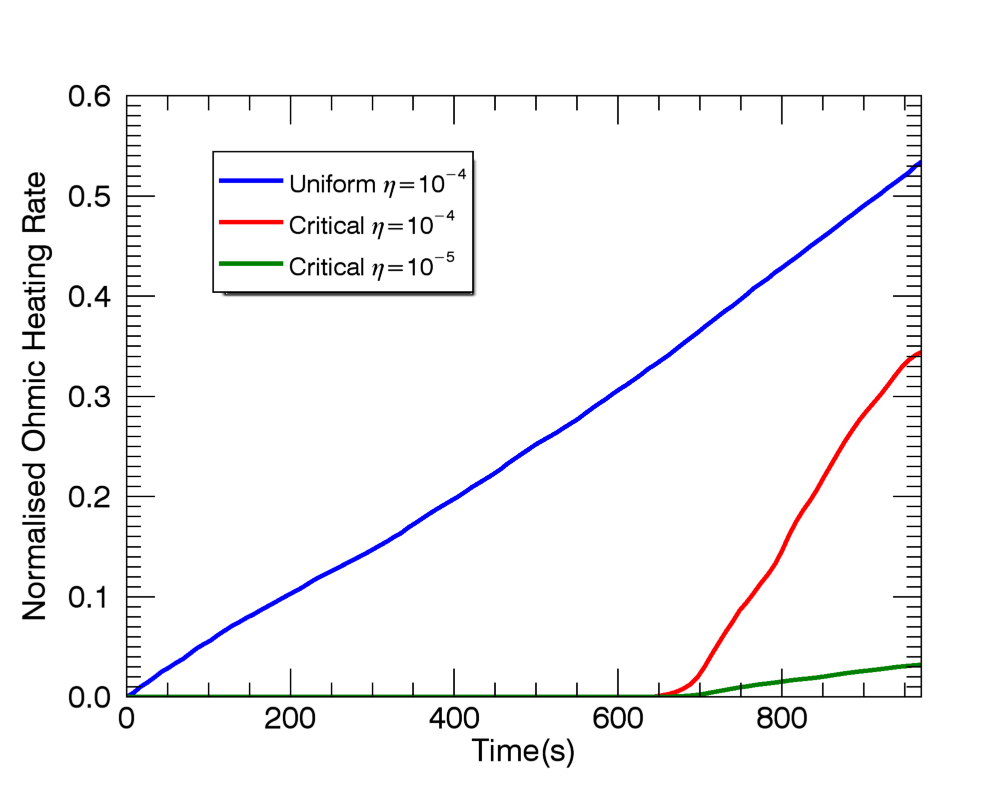}
  \caption{Ohmic heating rate at the loop apex for the $\left\{\eta=10^{-4}, \nu=10^{-4}\right\}$ case (blue), the $\left\{\eta=10^{-20}, \nu=10^{-20}, \eta^*=10^{-4} \right\}$ case (red) and the $\left\{\eta=10^{-20}, \nu=10^{-20}, \eta^*=10^{-5} \right\}$ case (green).}
  \label{crit_cur2}
\end{figure}

We conducted experiments with viscosities of $10^{-4}$ and $10^{-20}$, respectively. In both cases, we implemented a background $\eta = 10^{-20}$ and an enhanced $\eta^*$ of $10^{-4}$. In addition, we also considered a case with $\nu=10^{-4}, \eta=10^{-20}$ and $\eta^*=10^{-5}$. We chose a critical threshold that allowed the instability to form but then acted to increase the resistivity where current sheets formed in the shell region. The resistivity was only enhanced close to the loop apex and only at a small number of grid points within this sub-section of the domain. 

Until $\eta^*$ is triggered, the simulations are identical to those that do not consider the critical resistivity and even beyond this time, the large-scale behaviour remains unchanged. Indeed, cuts of the density profile at the loop apex (see Figure \ref{crit_cur1}) seem identical to the corresponding images in Figure \ref{Dens_Ev}. A more detailed analysis using contour plots of the density difference shown in Figure \ref{crit_cur1} finds only very minor changes around the KHI vortices in the shell region. Further, in contrast with the $\left\{\eta=10^{-4}, \nu=10^{-4}\right\}$ simulation, in which we found the KHI did not form, we find little evidence to suggest that the triggering of a critical resistivity leads to significant suppression of the instability once it has formed. This is unsurprising given the very small areas that are affected by the anomalously enhanced resistivity (Figure \ref{crit_cur1}).

The triggering of an $\eta^*$ that is many orders of magnitude larger than the background resistivity, inevitably leads to a dramatic increase in the Ohmic heating rate. We observe this in Figure \ref{crit_cur2} for the simulations that include the enhanced resistivity. Further, we also find that when $\eta^* = 10^{-4}$ is triggered, the rate of Ohmic heating observed at the loop apex exceeds the rate in the $\left\{\eta=10^{-4}, \nu=10^{-4}\right\}$ simulation in which the KHI is suppressed. This is remarkable given that $\eta=10^{-4}$ only at a few grid points in this case, as opposed to at all points in the domain. This supports the hypothesis that the formation of the KHI can enhance the deposition of heat beyond the level possible by the process of phase mixing alone.

\section{Discussion and conclusions}
 
The results that we have presented show that the effects of both resistivity and viscosity on the development of the Kelvin-Helmholtz instability in 3-D, numerical MHD simulations are significant. We modelled the evolution of a standing kink mode in a straight, density-enhanced magnetic flux tube. This wave exhibits rapid decay as energy is transferred through resonant absorption to azimuthal Alfv\'enic motions contained within the boundary region of the loop. This mode conversion is an ideal process and so does not depend on the dissipation coefficients. The resulting azimuthal Alfv\'{e}nic waves are associated with a large velocity shear that triggers the onset of the KHI. These waves exhibit much smaller length scales that are sensitive to both $\eta$ and $\nu$. Through the irreversible process of phase mixing, both of these dissipative plasma properties act to remove energy from the Alfv\'enic waves more efficiently than from the kink mode. Both the viscosity, and to a lesser extent, the resistivity contribute to the suppression of unstable, high wave-number, azimuthal modes and so act to restrict the growth of the instability.

The KHI is associated with the transfer of energy from shear flows in the boundary of the oscillating loop to vortical motions that cause the deformation of the loop's density profile. Tracking the extent of high density plasma acts as a proxy for both the instability onset time and the subsequent growth rate. Using this method, we were able to quantify the effects of both resistivity and viscosity on the development of the KHI. We note that whilst both transport coefficients suppress the instability, they have a non-symmetric effect with viscosity having a more important role in delaying the onset time. Viscosity acts directly to reduce the level of velocity shear, whereas resistivity has a more subtle effect by removing energy from the shell region. The effects of viscosity are greatest at the loop apex where the levels of vorticity are maximal. On the other hand, the currents in the domain are dominated by those close to the loop foot points and so the effects of resistivity are smallest near the loop apex. 

The growth of the KHI vortices is observed as an increase in the levels of vorticity exhibited at the loop apex. Analogously, the flows induced by the instability are compressive in nature and stress the magnetic field, generating significant currents within the loop's boundary (as observed in \cite{PatrickKHI}). Both the velocity and magnetic field gradients are subject to dissipation by viscosity and resistivity, respectively. Consequently, the formation of the instability is associated with an increase in thermal energy through enhanced viscous and Ohmic heating. Indeed, Figure \ref{Heat_fast} highlights that the onset of heating can occur earlier with lower dissipation coefficients as the KHI develops more efficiently. This also implies that even for the same value of resistivity, higher numerical resolution would lead to an increased heating rate as smaller length scales would be able to develop. Due to the small size of the initial perturbation, the magnitude of irreversible plasma heating is necessarily small in these experiments. In particular, it is certainly insignificant when compared to the increase in temperature in the loop boundary caused by the mixing of hot plasma from the loop exterior during the formation of the KHI. This effect is examined in detail in \citet{Fluting_modes}. Of course, in a constantly driven system more energy would be available for heating. For a more detailed analysis of the actual energy dissipation associated with kink modes in an inhomogeneous flux tube, we refer the reader to \citet{Pagano2017}, who also compare the obtained heating with coronal radiative losses. A comprehensive overview of observational signatures associated with resonant absorption, the formation of the KHI and phase mixing is given by \citet{Antolin2017}.

Although they are not well constrained, it is thought that typical values of coronal viscosity and resistivity are much smaller than the levels of $\eta$ and $\nu$ that we have found to suppress the formation of the KHI. However, our results may be directly applicable to transverse oscillations in chromospheric structures or in prominences. The temperature in such locations is typically much lower than the corona and so species are often only partially ionised. The presence of neutral atoms in the plasma enhances both the viscous and resistive effects. We direct the interested reader to \citet{Forteza2007, Soler2009, Soler2015, Khomenko2012} for further discussion on the effects of enhanced dissipation on chromospheric and prominence heating.

Due to the limits imposed by numerical dissipation, we cannot attain coronal values in large scale 3-D MHD models, such as the one presented here. The spatial resolution required to model coronal viscosity and resistivity accurately is too computationally expensive. Therefore, in order to reduce numerical energy loss, many coronal heating models implement artificially large dissipation to prevent the formation of unresolved small scales. This technique allows energy dissipation to be accounted for despite coarse spatial resolution. As we have shown, this may cause the suppression of the KHI, and impede investigations into the instability as a heating mechanism. The modified method of implementing a large, localised resistivity demonstrates that the Ohmic heating rate associated with the KHI may be more significant than that associated with phase mixing in an instability-suppressed regime.

The small length scales produced by the KHI are difficult to track accurately in 3-D numerical simulations. The turbulent plasma begins to exhibit variation on the grid scale and so numerical effects cause the loss of energy from the domain. Despite the turbulent flows (see bottom-right panel of Figure \ref{Dens_Ev}) necessarily being associated with enhanced dissipation, we are typically unable to monitor this energy conversion due to much of it being lost from the simulation. This has ramifications for investigating the KHI, and indeed other drivers of MHD turbulence, as coronal heating mechanisms. As the energy cascade reaches smaller and smaller scales, we inevitably reach a spatial resolution limit which means we cannot self-consistently model the heating of the turbulent plasma. Despite this, a plasma heating rate may be inferred by the extrapolation of results from a spatially resolved regime, however the reliability of this method remains unclear.

As discussed previously (see Introduction), there are many observations of transversely oscillating coronal loops that, according to our model, should be subject to the KHI. Hitherto, the lack of observational evidence of KHI formation in loops subject to the kink mode, may be expected due to the lack of spatial resolution of coronal imagers. The small length scales associated with the formation of a turbulent regime within a coronal loop are not expected to be directly observable by current technology. However, the results of forward modelling by \citet{Loop_compression}, suggest that observations of decay-less oscillations may be evidence of the spatially under-resolved Kelvin-Helmholtz instability.

The KHI may be more readily suppressed in the solar corona than is suggested here. It is possible that twist in the magnetic field or non-uniform loop sub-structure will prevent the formation of the instability. However, regimes with small values of twist in the magnetic field or a thin resonant layer may still be unstable to the KHI. Furthermore, \citet{Twist_OK} find that twisted field lines do not suppress the development of the instability. Meanwhile, larger values of twist are susceptible to the development of the kink instability and may induce other energy release mechanisms \citep{Kink_heating}. The unknown nature of coronal loops confounds the question of the applicability of this model further. Until the form and structure of loops are better understood, it is difficult to determine how the structure assumed in the model's set-up compares to conditions in the corona.

\vspace{1cm}

The authors would like to thank the referee for their helpful comments and suggestions. The authors would also like to thank Prof. P. J. Cargill for helpful discussions. The research leading to these results has received funding from the UK Science and Technology Facilities Council and the European Union Horizon 2020 research and innovation programme (grant agreement No. 647214). This work used the DIRAC 1, UKMHD Consortium machine at the University of St Andrews and the Darwin Data Analytic system at the University of Cambridge, operated by the University of Cambridge High Performance Computing Service on behalf of the STFC DiRAC HPC Facility (www.dirac.ac.uk). This equipment was funded by a BIS National E-infrastructure capital grant (ST/K001590/1), STFC capital grants ST/H008861/1 and ST/H00887X/1, and DiRAC Operations grant ST/K00333X/1. DiRAC is part of the National E-Infrastructure.

\bibliographystyle{aa}        
\bibliography{KHI}           

\end{document}